\documentstyle[aps]{revtex}
\newcommand{\be}{\begin{equation}}
\newcommand{\ee}{\end{equation}}
\newcommand{\ba}{\begin{eqnarray}}
\newcommand{\ea}{\end{eqnarray}}
\newcommand{\baa}{\begin{eqnarray*}}
\newcommand{\eaa}{\end{eqnarray*}}
\newcommand{\lab}[1]{\label{#1}}

\newcommand{\dis}{\displaystyle}

\begin{document}
\draft
{\pagestyle{empty}
\vskip 1.5cm

\title{Replica-exchange simulated tempering method for simulations
of frustrated systems}

\maketitle
\vskip 1.0cm

\centerline{Ayori Mitsutake$^{a,b,}$\footnote{\ \ e-mail: ayori@ims.ac.jp}
and Yuko Okamoto$^{a,b,}$\footnote{\ \ e-mail: okamotoy@ims.ac.jp}}
 
\vskip 1.0cm
  
\centerline{{\it $^a$Department of 
Theoretical Studies, Institute for Molecular Science}}
\centerline{{\it Okazaki, Aichi 444-8585, Japan}}
\centerline{and}
\centerline{{\it $^b$Department of Functional Molecular Science,
The Graduate University for Advanced Studies}}
\centerline{{\it Okazaki, Aichi 444-8585, Japan}}

\vfill

\begin{abstract}
 We propose a new
 method for the determination of the weight factor for the
simulated tempering method.
 In this method a short replica-exchange simulation is performed
 and the simulated tempering weight factor is obtained by the multiple-histogram
 reweighting techniques.
The new algorithm is particularly useful for studying frustrated systems with
rough energy landscape where the determination of the simulated tempering
weight factor by the usual iterative process becomes very difficult.
The effectiveness of the method is illustrated by taking an example
for protein folding.
 \end{abstract}

\vfill

}
\newpage

\section{Introduction}
In complex systems such as spin glasses and biopolymers,
it is very difficult to obtain accurate canonical distributions
at low temperatures by conventional Monte Carlo (MC) and
molecular dynamics (MD) simulation methods.
This is because simulations at low temperatures tend to get
trapped in one of huge number of local-minimum-energy states.
One way to overcome this multiple-minima
problem is to perform a simulation in a {\it generalized ensemble} where
each state is weighted by a non-Boltzmann probability
weight factor so that
a random walk in potential energy space may be realized.
(For a review of generalized-ensemble approach in the
protein folding problem, see, e.g., 
Ref.~\cite{RevHO}.)
The random walk allows the simulation to escape from any
energy barrier and to sample much wider phase space than
by conventional methods.
Monitoring the energy in a single simulation run, one can 
obtain not only
the global-minimum-energy state but also canonical ensemble
averages as functions of temperature by the single-histogram \cite{FS1}
and/or multiple-histogram \cite{FS2,WHAM} reweighting techniques.

Two of the most well-known generalized-ensemble algorithms are perhaps 
{\it multicanonical algorithm} (MUCA) \cite{MUCA1} and 
{\it simulated tempering} (ST) \cite{ST1,ST2}
(the latter method is also referred to as the 
{\it method of expanded ensemble} \cite{ST1}).
(For reviews, see, e.g., Refs.~\cite{MUCArev,STrev}).  
A simulation in MUCA performs a free
1D random walk in potential energy space.
This method and its generalizations have already been 
used in many applications in
protein systems (see, for instance,
Refs.~\cite{HO} -- \cite{MO3}).
A simulation in ST
performs a free 1D random walk in temperature space, which
in turn 
induces a random walk in potential energy space and
allows the simulation to escape from
states of energy local minima again.
ST has also been applied to protein folding
problem \cite{IRB1,HO96a,HO96b,IRB2}.
    
The generalized-ensemble algorithms
are powerful, but in the above two methods the probability
weight factors are not {\it a priori} known and have to be
determined by iterations of short trial simulations
\cite{MUCA1,ST1,ST2}.
This process can be non-trivial and very tedious for
complex systems with many local-minimum-energy states.

The {\it replica-exchange method} (REM) 
\cite{RE1,RE2} alleviates this difficulty.  (A similar method was 
independently 
developed earlier in Ref.~\cite{RE3}. 
REM is also referred to as {\it multiple Markov chain method} \cite{RE4}
and {\it parallel tempering} \cite{STrev}.) 
In this method, a number of
non-interacting copies of the original system (or replicas) 
at different temperatures are
simulated independently and
simultaneously by the conventional MC or MD methods. Every few steps,
pairs of replicas are exchanged with a specified transition
probability.
The weight factor is just the
product of Boltzmann factors, and so it is essentially known.
We have developed a molecular dynamics algorithm
for REM \cite{SO} (see, also, Ref.~\cite{H97}).  
We then developed a multidimensional REM \cite{SKO}.

However, REM also has a computational difficulty:
As the number of degrees of freedom of the system increases, 
the required number of replicas also increases, and so does
the required computation time. This is why we want to combine the
merits of MUCA and ST and those of REM
so that we can determine the weight factors for MUCA and ST
with ease and save the computation time greatly.
We have presented such an example; we developed 
a new method for the determination
of the multicanonical weight factor \cite{SO3,MSO}.  The method was 
referred to as
the {\it replica-exchange multicanonical algorithm} (REMUCA).
In REMUCA,
a short replica-exchange simulation is performed, and the multicanonical
weight factor is determined by
the multiple-histogram reweighting techniques \cite{FS2,WHAM}.

In this Letter we present another example of such a combination.
In the new method, which we refer to as the
{\it replica-exchange simulated tempering} (REST),
a short replica-exchange simulation is performed, and
the simulated tempering weight factor is determined by
the multiple-histogram reweighting techniques \cite{FS2,WHAM}. 
The effectiveness of the method is tested with a penta 
peptide, Met-enkephalin, in gas phase. 

\section{Methods}

We first briefly review the original {\it simulated tempering} (ST) 
method \cite{ST1,ST2}.  In this method temperature itself becomes a
dynamical variable, and both the configuration and the temperature are updated
during the simulation with a weight:
\begin{equation}
W_{{\rm ST}} (E;T) = e^{-\beta E + a(T)}~,
\label{eqn1}
\end{equation}
where $\beta = 1/k_B T$ ($k_B$ is the Boltzmann constant), $E$ is the
potential energy, and
the function $a(T)$ is chosen so that the probability distribution
of temperature is given by
\begin{equation}
P_{{\rm ST}}(T) = \int dE~ n(E)~ W_{{\rm ST}} (E;T) = 
\int dE~ n(E)~ e^{-\beta E + a(T)} = {\rm const}~,
\lab{eqn2}
\end{equation}
where $n(E)$ is the density of states.
Hence, in ST the {\it temperature} is sampled
uniformly. A free random walk in temperature space
is realized, which in turn 
induces a random walk in potential energy space and
allows the simulation to escape from
states of energy local minima.

In the numerical work we discretize the temperature in
$M$ different values, $T_m$ ($m=1, \cdots, M$); we can order the temperature
so that $T_1 < T_2 < \cdots < T_M$.  The lowest temperature 
$T_1$ should be sufficiently low so that the simulation can explore the
global-minimum-energy region, and
the highest temperature $T_M$ should be sufficiently high so that
no trapping in a local-minimum-energy state occurs.  The probability
weight factor in Eq.~(\ref{eqn1}) is then given by
\begin{equation}
W_{{\rm ST}}(E;T_m) = e^{-\beta_m E + a_m}~,
\label{eqn3}
\end{equation}
where $a_m=a(T_m)$ ($m=1, \cdots, M$).  
The parameters $a_m$ are not {\it a priori} known and have to be determined
by iterations of short simulations (see, e.g.,  Refs.~\cite{STrev,IRB1,HO96b} 
for details).
This process can be non-trivial and very difficult for complex systems.
Note that from Eqs.~(\ref{eqn2}) and (\ref{eqn3}) we have
\begin{equation}
e^{-a_m} \propto \int dE~ n(E)~ e^{- \beta_m E}~. 
\label{eqn4}
\end{equation}
The parameters $a_m$ are therefore ``dimensionless'' Helmholtz free energy
at temperature $T_m$
(i.e., the inverse temperature $\beta_m$ multiplied by 
the Helmholtz free energy).

Once the simulated tempering parameters $a_m$ are determined and 
the initial configuration and
the initial temperature $T_m$ are chosen,
a ST simulation is realized by alternately 
performing the following two steps \cite{ST1,ST2}:
\begin{enumerate}
\item A canonical MC or MD simulation at the fixed temperature $T_m$ 
is carried out for a certain MC or MD steps. 
\item The temperature $T_m$ is updated to the neighboring values
$T_{m \pm 1}$ with the configuration fixed.  The transition probability of
this temperature-updating
process is given by the Metropolis criterion (see Eq.~(\ref{eqn3})):
\begin{equation}
w(T_m \rightarrow T_{m \pm 1}) 
= \left\{
\begin{array}{ll}
 1~, & {\rm for} \ \Delta \le 0~, \cr
 \exp \left( - \Delta \right)~, & {\rm for} \ \Delta > 0~,
\end{array}
\right.
\label{eqn5}
\end{equation}
where
\begin{equation}
\Delta = \left(\beta_{m \pm 1} - \beta_m \right) E
- \left(a_{m \pm 1} - a_m \right)~.
\label{eqn6}
\end{equation}
\end{enumerate}
In the present work, we employ Monte Carlo algorithm
for Step 1.
After a long ST production run, one can obtain
canonical ensemble averages as functions of temperature by
the multiple-histogram reweighting techniques \cite{FS2,WHAM}
as described in detail below.

The {\it replica-exchange method} (REM)
was developed as an extension of
{\it simulated tempering} \cite{RE1} (thus it is also referred to as
{\it parallel tempering} \cite{STrev})
(see, e.g.,  Ref.~\cite{SO} for a detailed
description of the algorithm).
The system for REM consists of
$M$ {\it non-interacting} copies (or, replicas)
of the original system in the canonical ensemble
at $M$ different temperatures $T_m$ ($m=1, \cdots, M$).
Let $X = \left\{\cdots, x_m^{[i]}, \cdots \right\}$
stand for a state in this generalized ensemble.
Here, the superscript $i$ and the subscript $m$ in $x_m^{[i]}$
label the replica and the temperature, respectively.
The state $X$ is specified by the $M$ sets of
coordinates $q^{[i]}$ (and momenta $p^{[i]}$)
of the atoms in replica $i$ at
temperature $T_m$ ($i, m = 1, \cdots, M$):
\begin{equation}
x_m^{[i]} \equiv \left\{q^{[i]},p^{[i]}\right\}_m~.
\label{eqn6b}
\end{equation}
In Monte Carlo algorithm we need to take into account
only $q^{[i]}$.

A REM simulation 
is then realized by alternately performing the following two
steps \cite{RE1} --\cite{RE3}.
\begin{enumerate}
\item Each replica in canonical ensemble of the fixed temperature
is simulated $simultaneously$ and $independently$
for a certain MC or MD steps.
\item A pair of replicas,
say $i$ and $j$, which are
at neighboring temperatures, $T_m$ and $T_{m \pm 1}$, respectively,
are exchanged:
$X = \left\{\cdots, x_m^{[i]}, \cdots, x_{m \pm 1}^{[j]}, \cdots \right\}
\longrightarrow \
X^{\prime} = \left\{\cdots, x_m^{[j]}, \cdots, x_{m \pm 1}^{[i]},
\cdots \right\}$.
The transition probability of this replica-exchange process is given by
the Metropolis criterion:
\begin{equation}
w(X \rightarrow X^{\prime}) 
= \left\{
\begin{array}{ll}
 1~, & {\rm for} \ \Delta \le 0~, \cr
 \exp \left( - \Delta \right)~, & {\rm for} \ \Delta > 0~,
\end{array}
\right.
\label{eqn7}
\end{equation}
where
\begin{equation}
\Delta = \left(\beta_{m \pm 1} - \beta_m \right)
\left(E\left(q^{[i]}\right) - E\left(q^{[j]}\right)\right)~.
\label{eqn7b}
\end{equation}
Here, 
$E\left(q^{[i]}\right)$ and $E\left(q^{[j]}\right)$
are the potential energy of the $i$-th replica
and the $j$-th replica, respectively.
\end{enumerate}
In the present work we employ Monte Carlo algorithm
for Step 1.
A random walk in temperature space is
realized for each replica, which in turn induces a random
walk in potential energy space.  This alleviates the problem
of getting trapped in states of energy local minima.

The major advantage of REM over other generalized-ensemble
methods such as MUCA \cite{MUCA1} and ST \cite{ST1,ST2}
lies in the fact that the weight factor
is essentially {\it a priori} known (which is just a product
of Boltzmann factors),
whereas in the latter algorithms the determination of the
weight factors can be very tedious and time-consuming.
However, the number of required replicas (or temperatures)
for REM increases like $\sqrt N$ where $N$ is the number
of degrees of freedom of the system \cite{RE1}.
We will need a huge number of replicas and a large
amount of computation time to simulate
a complex system such as a real protein system,
while in MUCA or ST
only one replica (the original system itself) is simulated.
This led us to combine the merits of REM and ST
(the combination of REM and MUCA was
also realized \cite{SO3,MSO}).

We finally present the new method which we refer to as the 
{\it replica-exchange simulated tempering} (REST).  In this method 
we first perform a short REM simulation (with $M$ replicas)
to determine the simulated tempering
weight factor and then perform with this weight
factor a regular ST simulation with high statistics.
The first step is accomplished by the multiple-histogram reweighting
techniques \cite{FS2,WHAM} as follows.
Let $N_m(E)$ and $n_m$ be respectively
the potential-energy histogram and the total number of
samples obtained at temperature $T_m=1/k_B \beta_m$ of the REM run.
The density of states, $n(E)$,
is then given by \cite{FS2,WHAM}
\begin{equation}
n(E) = \frac{\dis{\sum_{m=1}^M g_m^{-1}~N_m(E)}}
{\dis{\sum_{m=1}^M n_{m}~g_m^{-1}~e^{f_m-\beta_m E}}}~,
\label{eqn8a}
\end{equation}
where
\begin{equation}
e^{-f_m} = \sum_{E} n(E) e^{-\beta_m E}~.
\label{eqn8b}
\end{equation}
Here, $g_m = 1 + 2 \tau_m$,
and $\tau_m$ is the integrated
autocorrelation time at temperature $T_m$.
Note that 
Eqs.~(\ref{eqn8a}) and
(\ref{eqn8b}) are solved self-consistently
by iteration to obtain
the density of states $n(E)$ and the dimensionless
Helmholtz free energy $f_m$.

Once the estimate of the dimensionless Helmholtz free energy $f_m$ are
obtained, the simulated tempering 
weight factor can be directly determined by using
Eq.~(\ref{eqn3}) where we set $a_m = f_m$ (compare Eqs.~(\ref{eqn4})
and (\ref{eqn8b})).
A long ST run is then performed with this
weight factor.  
Let $N_m(E)$ and $n_m$ be respectively
the potential-energy histogram and the total number of
samples obtained at temperature $T_m=1/k_B \beta_m$ from this
ST run.  The multiple-histogram
reweighting techniques of Eqs.~(\ref{eqn8a}) and (\ref{eqn8b}) can be used
again to obtain the best estimate of the density of states
$n(E)$.
The expectation value of a physical quantity $A$
at any temperature $T~(= 1/k_B \beta)$ is then calculated from
\begin{equation}
<A>_{T} \ = \frac{\dis{\sum_{E}A(E)~n(E)~e^{-\beta E}}}
{\dis{\sum_{E} n(E)~e^{-\beta E}}}~.
\label{eqn9}
\end{equation}

\section{Results and discussion}
We tested the effectiveness of the new algorithm for the system of
a penta-peptide, Met-enkephalin, in gas phase. This peptide has the
amino-acid sequence, Tyr-Gly-Gly-Phe-Met.  The backbone was terminated 
by a neutral NH$_2$-~group and a neutral~-COOH group 
at the N-terminus and
at the C-terminus, respectively, as in the previous MC works of Met-enkephalin.
The potential energy function $E_{{\rm tot}}$ (in kcal/mol)
is given by the sum of
the electrostatic term $E_{\rm C}$, 12-6 Lennard-Jones term
$E_{{\rm LJ}}$, and hydrogen-bond term $E_{{\rm HB}}$ for all pairs of atoms in
the molecule together with the torsion term $E_{{\rm TOR}}$ for all torsion
angles.  The parameters
in the energy function as well as the molecular geometry
were taken from ECEPP/2 \cite{EC3}.
The computer code KONF90 \cite{KONF} was used, and
MC simulations 
based on the replica-exchange simulated tempering (REST)
were performed.
The dielectric
constant $\epsilon$ was set equal to 2 as in the previous works.
The peptide-bond dihedral angles $\omega$ were fixed at the value
180$^{\circ}$ for simplicity.
With this choice of parameters our results can be directly compared with Refs.
\cite{HO,MHO}.

The remaining dihedral angles constitute the variables to be
updated in the MC simulations:
$\phi_{k}$ and $\psi_{k}$ in the main chain ($k=1, \cdots, 5$) and
$\chi_{k}^{j}$ in the side chains
(there are 3 $\chi$'s for Tyr, 2 $\chi$'s for Phe, and 4 $\chi$'s for Met).
For Met-enkephalin the number of degrees of freedom is thus 19.
One MC sweep consists
of updating all these 19 angles once with
Metropolis evaluation for each update.
The simulations (of all replicas)
were started from randomly generated conformations.

In Table I we summarize the parameters of the simulations that
were performed in the present work. 
As described in the previous section, REST consists of 
two simulations: a short REM simulation (from which the 
dimensionless Helmholtz free energy, or the simulated tempering
weight factor, is determined) and a subsequent
ST production run.
The former simulation is referred to as REM1 and the latter
as ST1 in Table I.
In REM1 there exist
8 replicas with 8 different temperatures ($M=8$), ranging from
50 K to 1000 K as listed in Table I (i.e., $T_1 = 50$ K
and $T_M = T_{8} = 1000$ K).  
The same set of temperatures were also used in ST1.
The temperatures were distributed exponentially between
$T_1$ and $T_M$, following the optimal
distribution found in the previous simulated annealing
schedule \cite{KONF}, simulated tempering run \cite{HO96b},
and replica-exchange simulation \cite{SO}. 
Before taking the data in REM1, we made a REM simulation of
10,000 MC sweeps for thermalization.  We then performed
a REM simulation of $5 \times 10^4$ MC sweeps to obtain
the weight factor for simulated tempering.
After estimating the weight factor, we made a ST production
run of $10^6$ MC sweeps (ST1), which followed
additional 1,000 MC sweeps for equilibration.
In REM1 and ST1, a replica exchange and a temperature update,
respectively, were tried every 10 MC sweeps.
Data were collected at each MC sweep in both REM1 and ST1.

We first check whether the replica-exchange simulation
of REM1 indeed performed properly.
For an optimal performance of REM the acceptance ratios 
of replica exchange should be
sufficiently uniform and large (say, $> 10$ \%).
In Table II we list these quantities.
It is clear that both points are met in the sense that
they are of the same order (the values vary 
between 10 \% and 40 \%). 
Moreover, in Fig.~1 the canonical 
probability distributions obtained
at the chosen 8 temperatures from REM1
are shown.  We see that there
are enough overlaps between all pairs of neighboring
distributions, indicating
that there will be sufficient numbers of replica exchange
between pairs of replicas (see Table II).
We did observe random walks in 
potential energy space between low energies and high energies.

After REM1, we obtained the dimensionless Helmholtz free
energy at the eight temperatures, $f_m$ ($m=1, \cdots, 8$), 
by the multiple-histogram reweighting techniques \cite{FS2,WHAM} (namely, by
solving Eqs.~(\ref{eqn8a}) and (\ref{eqn8b}) self-consistently). 
This gives the simulated tempering weight factor of 
Eq.~(\ref{eqn3}) with $a_m = f_m$.
We remark that for biomolecular systems the integrated 
autocorrelation times, $\tau_m$, in the reweighting formulae
can safely be set to be a constant \cite{WHAM}, and we
do so throughout the analyses in this section.

After determining the simulated tempering weight factor,
we carried out a long ST simulation 
for data collection (ST1 in Table I).
In Fig. 2 the time series of temperature and
potential energy from ST1 are plotted.
In Fig.~2(a) we observe a random walk in
the temperature space between the lowest and highest
temperatures.  In Fig.~2(b) the corresponding random
walk of the total potential energy 
between low and high energies is observed.
Note that there is a strong correlation between the behaviors
in Figs.~2(a) and 2(b), as there should.
It is known from our previous works that the 
global-minimum-energy conformation for Met-enkephalin 
has the potential energy value
of $-12.2$ kcal/mol \cite{HO,MHO}.
Hence, the random walk in Fig.~2(b) indeed visited
the global-minimum region many times.  It also visited
high energy regions, judging from the fact that
the average potential energy
is around 15 kcal/mol at $T = 1000$ K \cite{HO,MHO}
(see Fig.~4 below).

For an optimal performance of ST, the acceptance ratios 
of temperature update should be
sufficiently uniform and large.
In Table III we list these quantities.
It is clear that both points are met (the values vary 
between 26 \% and 57 \%); 
we find that the present ST run (ST1)
indeed properly performed.
We remark that the acceptance ratios in Table III are
significantly larger and more uniform than those in Table II, suggesting
that ST runs can sample
the configurational space more effectively
than REM runs, provided the optimal weight factor is obtained.

In the previous
works of ST simulations of Met-enkephalin in gas phase
\cite{HO96a,HO96b}, at least several
iterations of trial simulations were required for the simulated
tempering weight determination.
We emphasize that in the present case of REST (REM1), only
one simulation was necessary to determine the optimal 
simulated tempering
weight factor that can cover the energy region corresponding to
temperatures between 50 K and 1000 K.

In Fig. 3 we plot the simulated tempering parameters $f_m$
($m=1, \cdots, 8$) and the dimensionless Helmholtz free energy
(inverse temperature multiplied by Helmholtz free energy)
$f(T)$ as a function of temperature $T$.  The former quantities,
$f_m$, were estimated by the multiple-histogram reweighting
techniques, using the results of the first REM run (REM1).
The latter quantity, $f(T)$, was calculated by the
multiple-histogram reweighting techniques, using the results
of the final ST production run (ST1).  Namely, the density
of states, $n(E)$, was obtained by solving
Eqs.~(\ref{eqn8a}) and (\ref{eqn8b}) by iteration.
The function $f(T)$ was then calculated by taking a summation
of $n(E) \exp(-\beta E)$ over $E$ for each value of $T$
(i.e., replace $f_m$ and $\beta_m$ in Eq.~(\ref{eqn8b})
by $f(T)$ and $\beta$, respectively).
The results agree very well with each other, implying that
the simulated tempering parameters, $f_m$, that were obtained
from a short REM run are already quite accurate.

To check the validity of the canonical-ensemble expectation values
calculated by the new method, in Fig. 4 we
compare the average potential energy as a function of temperature
obtained from ST1 with that obtained from the previous
MUCA simulation \cite{MHO}. 
The results for ST1 were obtained by the multiple-histogram 
method \cite{FS2,WHAM}
(see Eqs.~(\ref{eqn8a}), (\ref{eqn8b}), and (\ref{eqn9})), 
whereas the single-histogram method \cite{FS1}
was used for the results of the MUCA simulation. We can see a perfect 
coincidence of the average values in Fig. 4. 

\section{Conclusions}
In this Letter we have proposed a new algorithm for configurational
sampling of frustrated systems with rough energy landscape.
In this method, which we refer to as replica-exchange simulated
tempering (REST), the simulated tempering weight factor is
determined from
the results of a short replica-exchange simulation, and then a regular
simulated tempering production run is made with this weight.
The formulation of the method is simple and straightforward, but
the numerical improvement is great, because the weight factor
determination for simulated tempering becomes very difficult
by the usual iterative process for complex systems.
The new method was tested with 
the system of a small peptide in gas phase. 
The simulated tempering weight factor was indeed obtained by 
a single, short replica-exchange simulation.

\noindent
{\bf Acknowledgements}

The authors are grateful to Dr. Y. Sugita for useful discussions.
Our simulations were performed on the Hitachi and other computers at the
Research Center for Computational Science, Okazaki National Research
Institutes.
This work is supported, in part, by a grant from the Research
for the Future Program of the Japan Society for the Promotion of
Science (JSPS-RFTF98P01101).


%
%
%
%
\begin{table}
 \caption{Summary of parameters in REST simulations.}
 \begin{center}
 \begin{tabular}{cccc}
   Run     & No. of replicas & Temperature, $T_m$ (K) 
   & MC sweeps \\
   \hline
   REM1    & 8  & 50, 77, 118, 181, 277, 425, 652, 1000 & $5 \times 10^4$ \\
   ST1    & 1  & 50, 77, 118, 181, 277, 425, 652, 1000 & $1 \times 10^6$ \\
  \end{tabular}
 \end{center}
 \label{Tab1}
\end{table}

\begin{table}
 \caption{Acceptance ratios of replica exchange in REM1.}
 \begin{center}
 \begin{tabular}{cc}
   Pair of temperatures (K) & Acceptance ratio \\
   \hline
    50~~  $\longleftrightarrow$   ~~77 & 0.30 \\
    77~~  $\longleftrightarrow$   ~~118 & 0.27 \\
    118~~  $\longleftrightarrow$   ~~181 & 0.22 \\
    181~~  $\longleftrightarrow$   ~~277 & 0.17 \\
    277~~  $\longleftrightarrow$   ~~425 & 0.10 \\
    425~~  $\longleftrightarrow$   ~~652 & 0.27 \\
    652~~  $\longleftrightarrow$   ~~1000 & 0.40 \\
  \end{tabular}
 \end{center}
 \label{Tab2}
\end{table}
  
\begin{table}
 \caption{Acceptance ratios of temperature update in ST1.}
 \begin{center}
 \begin{tabular}{cc}
   Pair of temperatures (K) & Acceptance ratio \\
   \hline
    50~~  $\longrightarrow$   ~~77 & 0.47 \\
    77~~  $\longrightarrow$   ~~50 & 0.47 \\
    77~~  $\longrightarrow$   ~~118 & 0.43 \\
    118~~  $\longrightarrow$   ~~77 & 0.43 \\
    118~~  $\longrightarrow$   ~~181 & 0.37 \\
    181~~  $\longrightarrow$   ~~118 & 0.42 \\
    181~~  $\longrightarrow$   ~~277 & 0.29 \\
    277~~  $\longrightarrow$   ~~181 & 0.29 \\
    277~~  $\longrightarrow$   ~~425 & 0.30 \\
    425~~  $\longrightarrow$   ~~277 & 0.26 \\
    425~~  $\longrightarrow$   ~~652 & 0.43 \\
    652~~  $\longrightarrow$   ~~425 & 0.42 \\
    652~~  $\longrightarrow$   ~~1000 & 0.57 \\
    1000~~  $\longrightarrow$   ~~652 & 0.56 \\
  \end{tabular}
 \end{center}
 \label{Tab3}
\end{table}

\newpage

\centerline{\bf Figure Captions}

\begin{itemize}
 \item Fig.~1. Probability distribution of potential energy
               obtained at the eight temperatures
               from REM1 (see Table I for the parameters
               of the simulation).
               The left-most distribution corresponds to the lowest
               temperature ($T_1 = 50$ K), and the right-most
               to the highest one ($T_8 = 1000$ K).
 \item Fig.~2. Time series of temperature (a) and potential energy (b)
       in ST1 (see Table I for the parameters
               of the simulation).
 \item Fig.~3. Dimensionless Helmholtz free energy as a function of
       temperature $T$ (K).
        The dotted curve is the result from the simulated tempering
        production run (ST1).  The crosses are the 
        $f_m$ ($m=1, \cdots, 8$)
        that were determined 
        from the short preliminary replica-exchange run
        (REM1).  Both results were normalized so that the values
        at $T=50$ K agree with each other.
 \item Fig.~4. The average potential energy as a function of temperature.
       The solid curve was obtained by the multiple-histogram
       reweighting techniques from the results of ST1.
       The crosses were obtained by the single-histogram reweighting
       techniques from the results of the previous multicanonical
       MC simulation \cite{MHO}.

\end{itemize}
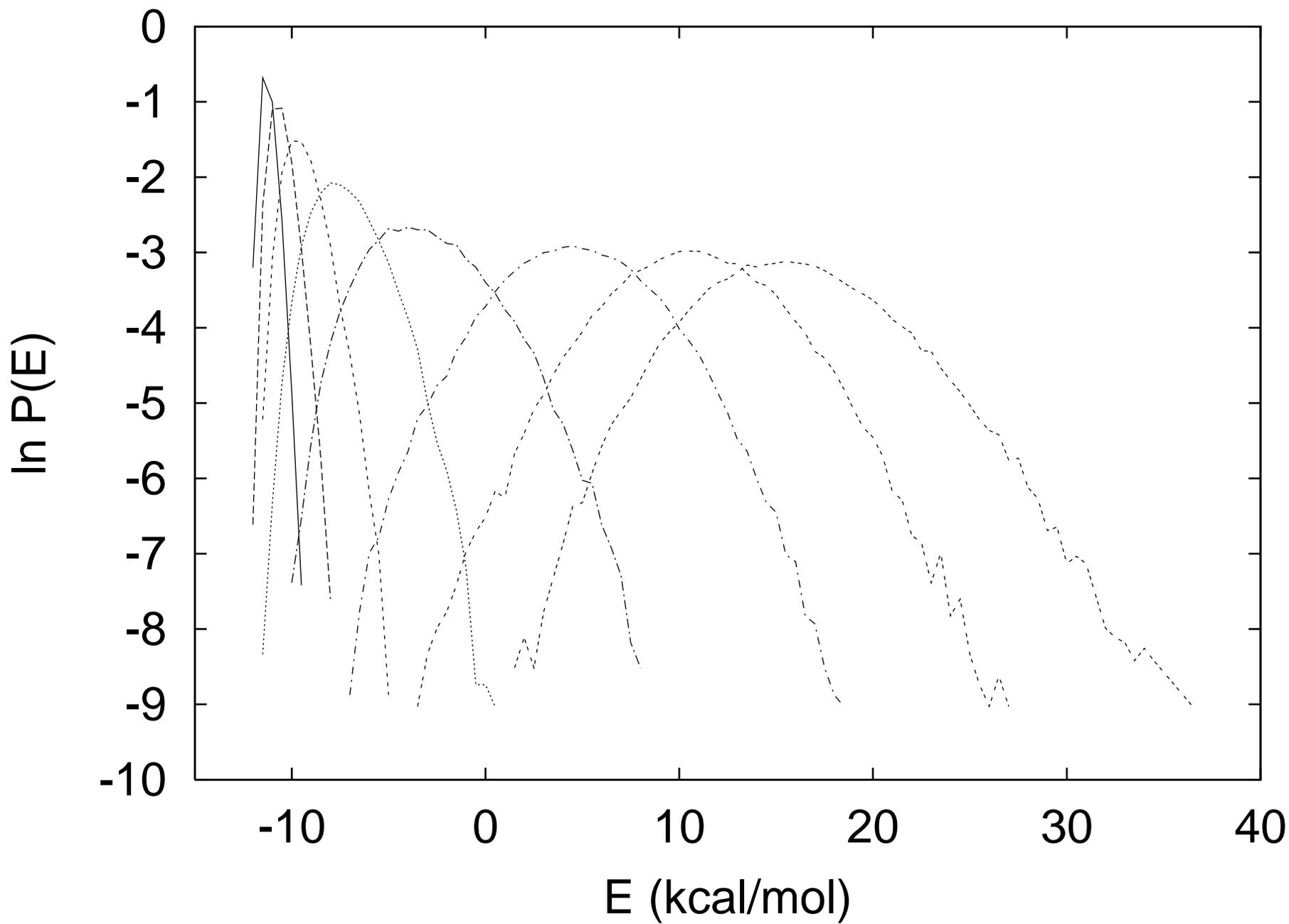
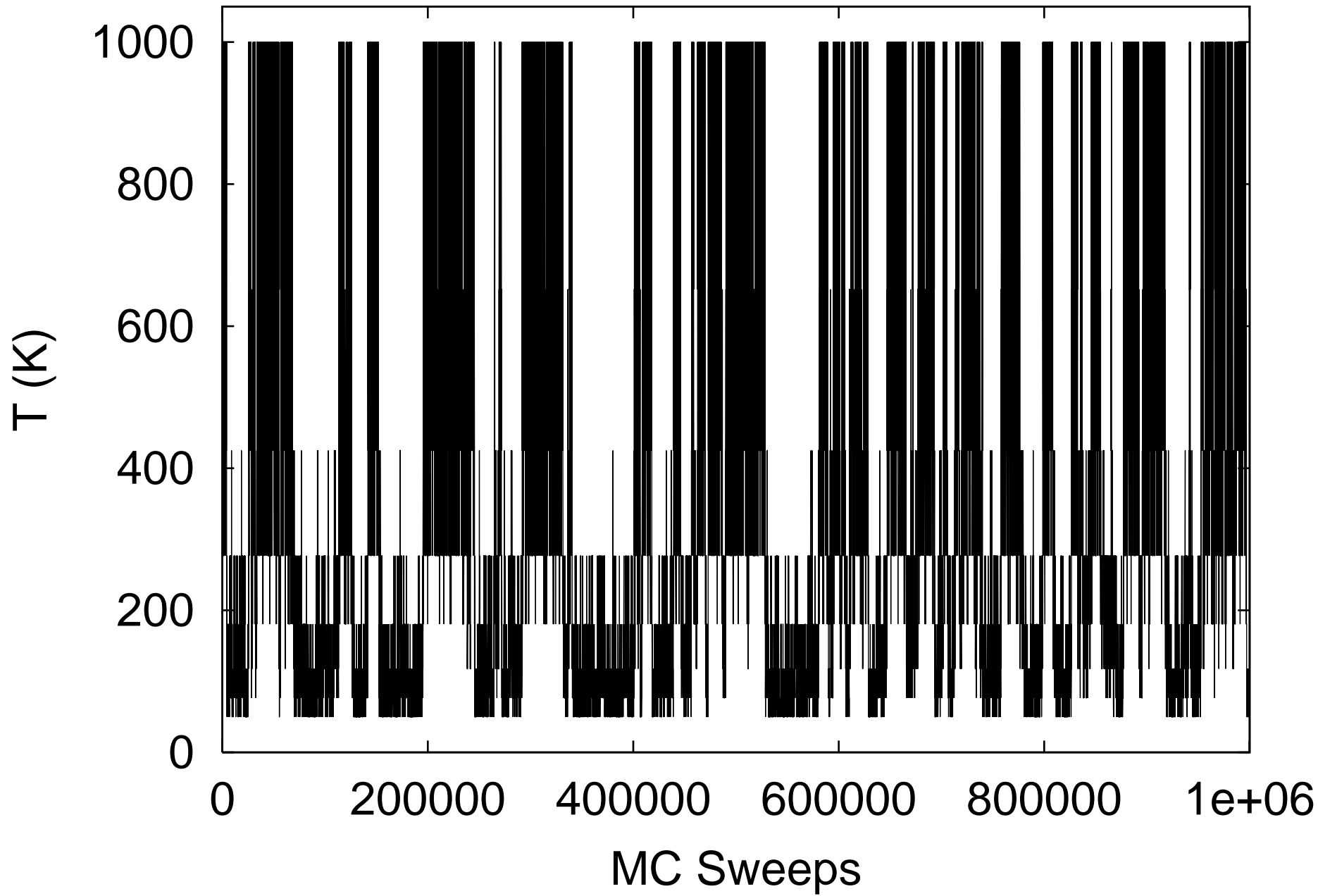
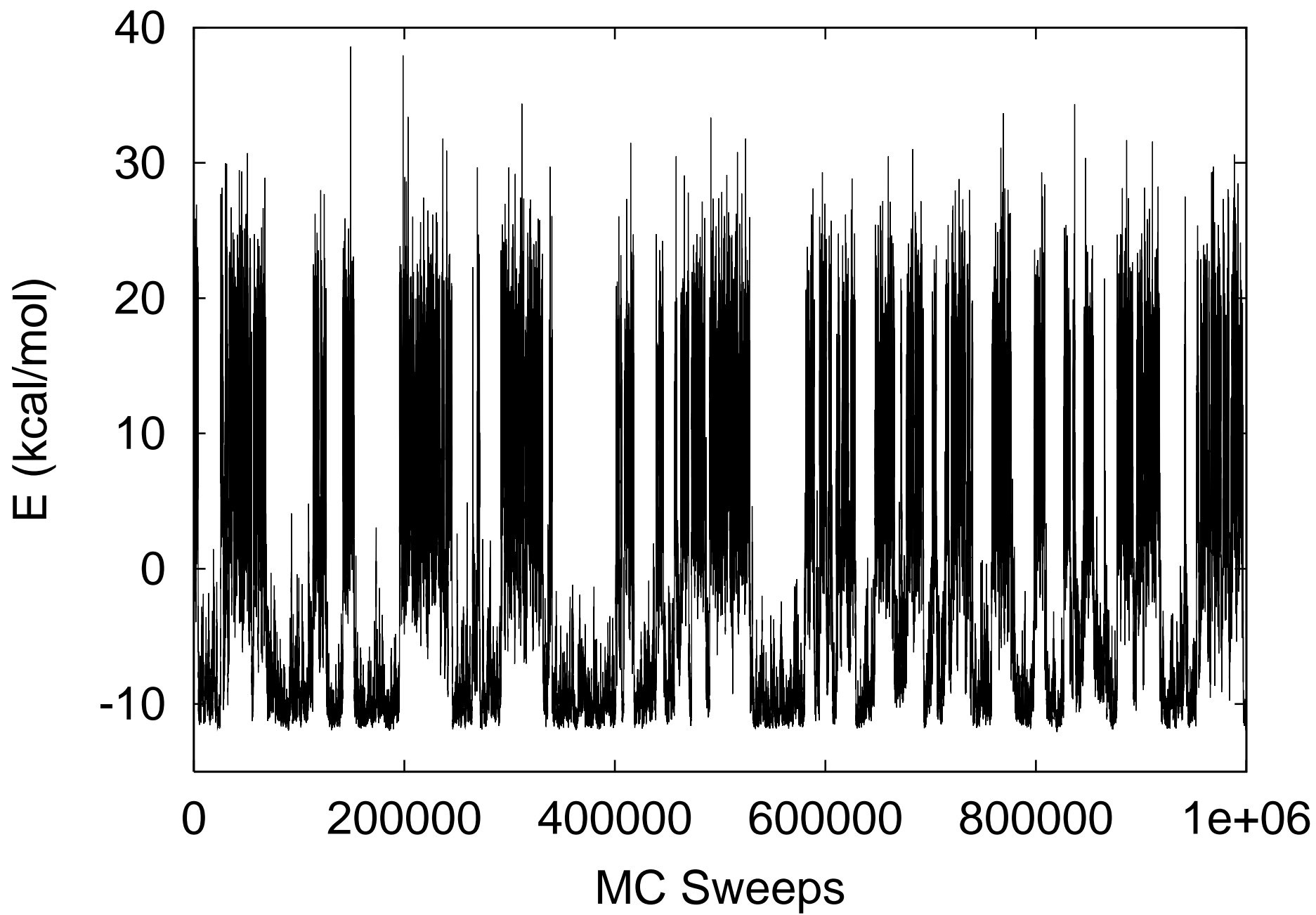
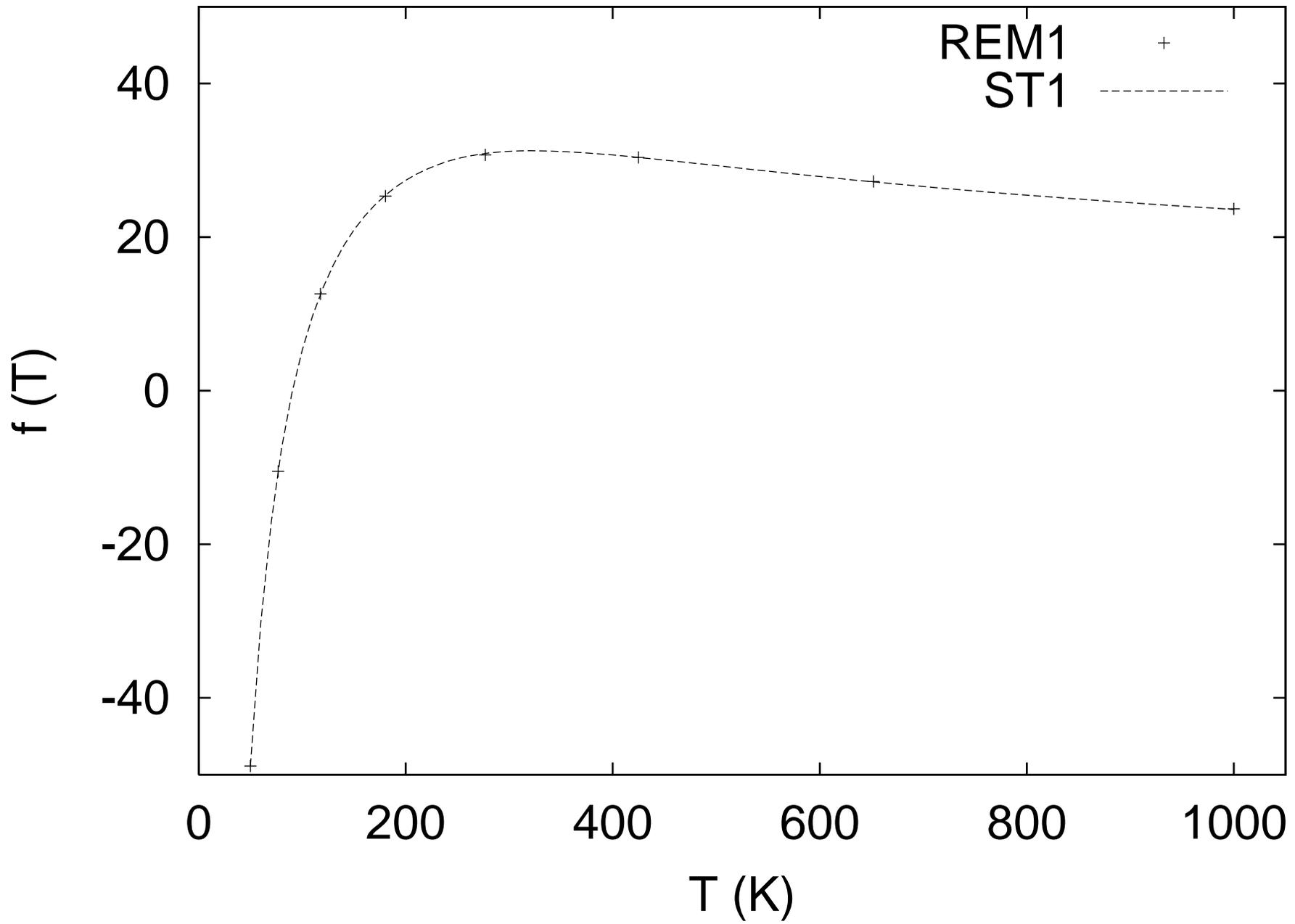
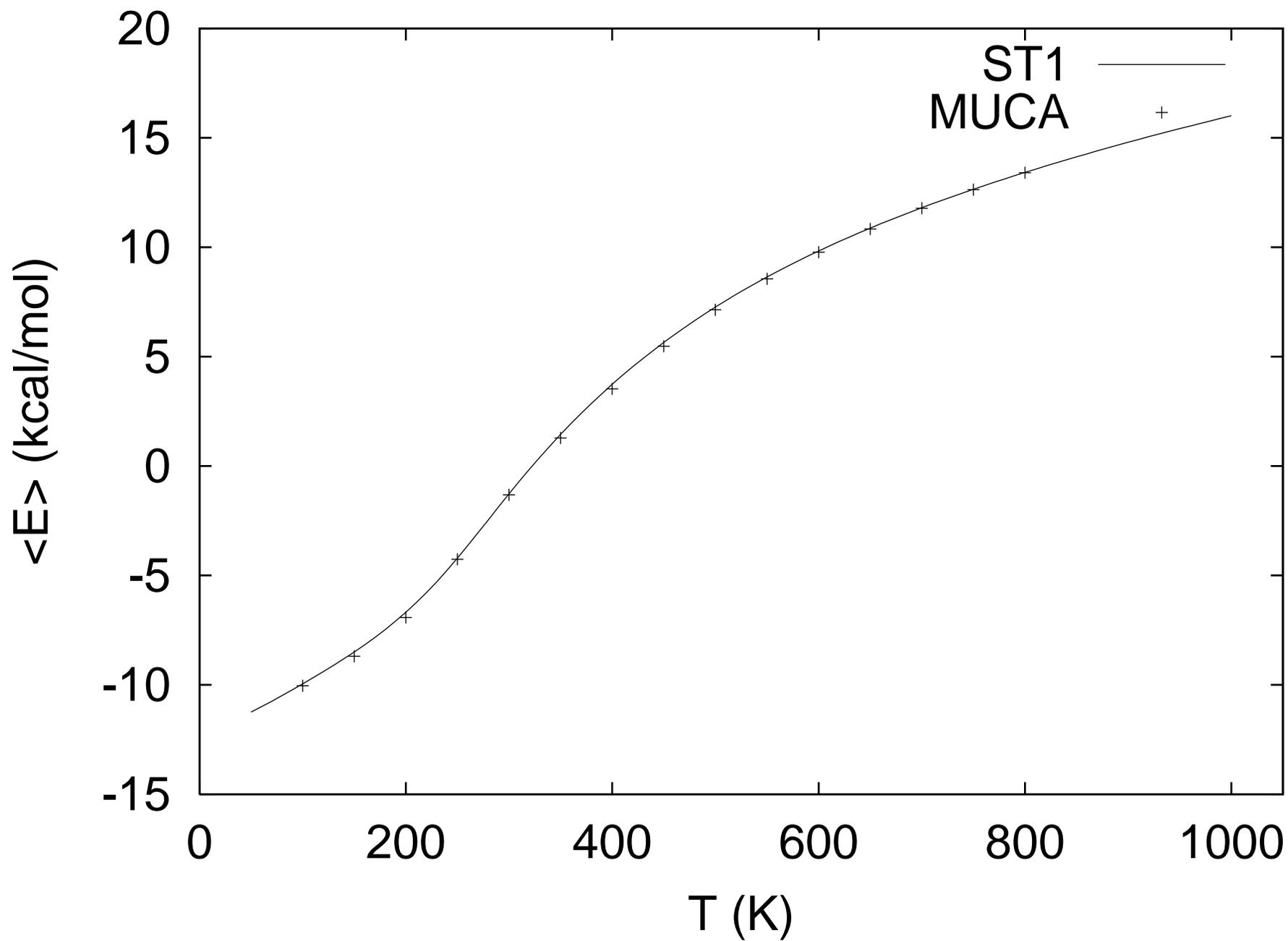
\end{document}